\documentclass[20pt]{article}
\usepackage{amsmath}
\usepackage{amssymb}
\usepackage{latexsym}
\usepackage{amsmath}
\usepackage{lineno}
\usepackage{graphicx}
\usepackage{float}
\usepackage{subfigure}
\usepackage{geometry}
\usepackage{graphics}
\usepackage{subfigure}
\usepackage{graphicx}
\usepackage{multirow}
\usepackage{booktabs}
\usepackage{hyperref}
\usepackage{cite}

\makeatletter
\newcommand{\be}{\begin{equation}}
\newcommand{\ee}{\end{equation}}

\begin{document}
\begin{center}
\large {\bf Rainbow gravity corrections to the entropic force}
\end{center}
\linenumbers
\begin{center}
Zhong-Wen Feng $^{1,2}$
 $\footnote{E-mail:  zwfengphy@163.com}$
Shu-Zheng Yang $^{1,2}$
$\footnote{E-mail:  szyangcwnu@126.com}$
\end{center}

\begin{center}
\textit{1.Physics and Space Science College, China West Normal University, Nanchong, 637009, China
2.Department of Astronomy, China West Normal University, Nanchong, 637009, China}
\end{center}

\noindent
{\bf Abstract:} The entropic force attracts a lot of interest for its multifunctional properties. For instance, Einstein's field equation, Newton's law of gravitation and the Friedmann equation can be derived from the entropic force. In this paper, utilizing a new kind of rainbow gravity model that was proposed by Magueijo and Smolin, we explore the quantum gravity corrections to the entropic force. First, we derive the modified thermodynamics of a rainbow black hole via its surface gravity. Then, according to Verlinde's theory, the quantum corrections to the entropic force are obtained. The result shows that the modified entropic force is related not only to the properties of the black hole but also the Planck length $\ell_p$, and the rainbow parameter  $\gamma$. Furthermore, based on the rainbow gravity corrected entropic force, the modified Einstein's field equation and the modified Friedmann equation are also derived.

\section{Introduction}
\label{Int}
Prior research generally confirms that black holes have thermal properties \cite{ch1,ch2,ch3,ch4}. This idea started out with an analogy connecting the laws of thermodynamics and those of gravity. In 1973, Bekenstein proved that the entropy of a black hole $S$  is proportional to its horizon area  $A$, that is  $S = {{Ak_B c^3 } \mathord{\left/ {\vphantom {{Ak_B c^3 } {4\hbar ^2 G}}} \right. \kern-\nulldelimiterspace} {4\hbar ^2 G}}$, where  $\hbar$ is the Planck constant, $k_B$ is the Boltzmann constant, and  $G$  is the Newton's gravitational constant, respectively \cite{ch1,ch2}. Subsequently,  inspired by Bekenstein's theory of entropy, Hawking first showed that the temperature of Schwarzschild black hole is  $T = {\kappa  \mathord{\left/ {\vphantom {\kappa  {2\pi }}} \right. \kern-\nulldelimiterspace} {2\pi }}$ with surface gravity  $\kappa$ \cite{ch3} . After years of continuous efforts, the authors in Ref.~\cite{ch8} established the laws of black hole thermodynamics, which imply that the thermodynamics of black holes have profound connections with gravity.

In the past twenty years, several theories have been proposed to study the deeper-seated relation between gravity and thermodynamics \cite{ch9,ch10,ch10+,ch11,ch12,ch13}. Among these perspectives, the entropic force approach, which was introduced by Verlinde, has attracted a lot of attention. In Ref.~\cite{ch13}, Verlinde claimed that gravity is not a fundamental force; instead, it would be explained as an entropic force, which arises from changes in the information associated with the positions of material bodies. This idea turns out to be very powerful. First, with the help of the entropic force approach and the Unruh temperature, one can obtained Newton's second law. Second, Newton's law of gravitation can be derived by incorporating the entropic force approach together with the holographic principle and the equipartition law of energy. Third, and most importantly, the entropic force approach results in Einstein's equation. Subsequently, according to this new proposal of gravity, many relevant works appeared. In Refs.~ \cite{ch15,ch16,ch17,ch18,ch18x}, the authors derived the Friedmann equation of a Friedmann-Robertson-Walker (FRW) universe via the entropic force approach. In Ref.~\cite{ch19}, the holographic dark energy has been investigated in the context of the entropic force approach. Nozari, Pedram and Molkara used the entropic force approach to study the Hamiltonians of the quantum systems in quasi-space \cite{ch20}. Shi, Hu and Zhao noted that entanglement entropy in loop quantum gravity can be used to derive an entropic force \cite{ch21}. Moreover, it should be noted that the effect of quantum gravity can influence the entropic force. Based on the entropic force approach,  Arag{\~a}o and Silva obtained Newtonian gravity in loop quantum gravity \cite{ch22}. In addition, the generalized uncertainty principle (GUP) corrected entropic force has been investigated in Refs.~\cite{ch23,ch24,ch25}.

On the other hand, it is well known that rainbow gravity (RG) is also motivated by quantum gravity \cite{ch31}. RG originates from the modified dispersion relation (MDR). In many quantum gravity theories, it is found that the Planck length is variable, which conflicts with previous conclusions \cite{ch316a,ch316b,ch316c}. To resolve this contradiction, the standard energy-momentum dispersion relation must be changed to a MDR. Interestingly, the MDR can modify special relativity. Currently, special relativity is called double special relativity (DSR) since it has two invariants: the velocity of light and the Planck length. \cite{ch316d}. In 2004, Magueijo and Smolin generalized DSR to curved spacetime and obtained RG. In RG, the spacetime background is related to the energy of a test particle. Therefore, the background of this spacetime can be described by a family of metrics (rainbow metrics) parameterized by $E \ell _p$. In Refs.~\cite{ch25+,ch26+}, Hendi \emph{et al.} studied the RG-corrected thermodynamics of black holes. Seiedgarr used the RG model that was introduced by Amelino-Camelia \emph{et al}. in Refs.~\cite{ch27,ch28} to study the modified  entropy-area law \cite{ch29}. His work showed that the modifications are related to the rainbow functions. As a matter of fact, there are many works demonstrating that the RG model is not unique. According to different phenomenological motivations, a series of RG models can be obtained. Based on the varying speed of light theory, which is a model that can explain the cosmological horizon, flatness, and cosmological constant problems of Big-Bang cosmology, Magueijo and Smolin proposed a kind of MDR
\begin{eqnarray}
\label{eq0}
{{E^2 } \mathord{\left/ {\vphantom {{E^2 } {\left( {1 - \gamma E\ell _p } \right)}}} \right. \kern-\nulldelimiterspace} {\left( {1 - \gamma E\ell _p } \right)}}^2  - p^2  = m^2,
\end{eqnarray}
where $\gamma$, $\ell _p$  and $m$ represent the rainbow parameter, the Planck length and the mass of the test particle, respectively. Obviously,  Eq.~(\ref{eq0}) indicates that spacetime has an energy-dependent velocity $c\left( E \right) = {{dE} \mathord{\left/ {\vphantom {{dE} {dp}}} \right. \kern-\nulldelimiterspace} {dp}} = 1 - E\ell _p$. Comparing it with the general form of MDR, namely, $E^2 f^2 \left( {E\ell _p } \right) - p^2 g^2 \left( {E\ell _p } \right) = m^2$, one can easily obtain the a new model of RG, whose rainbow functions are expressed as follows\cite{ch29+,ch29++}
\begin{eqnarray}
\label{eq1}
\begin{array}{*{20}c}
   {f\left( {E\ell _p } \right) = {1 \mathord{\left/
 {\vphantom {1 {\left( {1 - \gamma E\ell _p } \right)}}} \right.
 \kern-\nulldelimiterspace} {\left( {1 - \gamma E\ell _p } \right)}},} & {g\left( {E\ell _p } \right) = 1,}  \\
\end{array}
\end{eqnarray}
 It should be mentioned the rainbow functions are responsible for the modification of the energy-momentum relation in the ultraviolet regime. Besides, Eq.~(\ref{eq1}) satisfies
\begin{eqnarray}
\label{eq1+}
\begin{array}{*{20}c}
   {\mathop {\lim }\limits_{E\ell _p  \to 0} f\left( {E\ell _p } \right) = 1,} & {\mathop {\lim }\limits_{E\ell _p  \to 0} g\left( {E\ell _p } \right) = 1}.  \\
\end{array}
\end{eqnarray}
Using Eq.~(\ref{eq1}), Zhao and Liu calculated the modified thermodynamics of black objects \cite{ch30}. The results showed that the effect of rainbow gravity leads to new mass-entropy relations. In addition to that, we showed that Eq.~(\ref{eq1}) can influence the phase transition of black holes \cite{ch30+}. Inspired by those works, it is interesting to investigate the RG corrected entropic force. Therefore, in this paper, using the rainbow functions that were introduced by Magueijo and Smolin, we calculate the modified Hawking temperature and entropy of a black hole. Then, extending the work of Verlinde, the RG-corrected entropic force is yielded. Furthermore, according to the modified entropic force, the deformed Einstein's field equation and the deformed Friedmann equation are derived.

The organization of the paper is as follows. In Section~\ref{sec2}, we calculate the modified Hawking temperature and entropy via the surface gravity from RG. In Section~\ref{sec3}, the modified entropic force is derived. Section~\ref{sec4} and Section~\ref{sec5} are devoted to the derivation of the deformed Einstein's field equation and the deformed Friedmann equation, respectively. Finally, the results are briefly discussed in Section~\ref{sec6}.

Throughout the paper, the natural units of $k_B  = c = \hbar  = 1$ are used.

\section{The impact of rainbow gravity on black hole thermodynamics}
\label{sec2}
Generally, in order to obtain the modified metric of black holes in rainbow gravity, one can simply make the replacements ${\rm d}t \to {{{\rm d}t} \mathord{\left/  {\vphantom {{{\rm d}t} {f\left( {E\ell _p } \right)}}} \right. \kern-\nulldelimiterspace} {f\left( {E\ell _p } \right)}}$ for time coordinates and ${\rm d}x^i  \to {{{\rm d}x^i } \mathord{\left/ {\vphantom {{{\rm d}x^i } g}} \right. \kern-\nulldelimiterspace} g}\left( {E\ell _p } \right)$  for all spatial coordinates. Therefore, the metric of a static spherically symmetric black hole without charge in the framework of RG is
\begin{align}
\label{eq2}
{ d}{{s}}^2 & =  - \frac{{1 - \left( {{{2GM} \mathord{\left/  {\vphantom {{2GM} r}} \right.  \kern-\nulldelimiterspace} r}} \right)}}{{f^2 \left( {E\ell _p } \right)}}{d}{{t}}^2  + \frac{{{ d}{r}^2 }}{{\left[ {1 - \left( {{{2GM} \mathord{\left/  {\vphantom {{2GM} r}} \right.  \kern-\nulldelimiterspace} r}} \right)} \right]g^2 \left( {E\ell _p } \right)}} + \frac{{r^2 }}{{g^2 \left( {E\ell _p } \right)}}{\rm d}\Omega ^2 ,
\end{align}
where  $M$,  $G$, and ${\rm d}\Omega ^2$ are the mass of a static spherically symmetric black hole, Newton's gravitational constant, and line elements of hypersurfaces, respectively \cite{ch31}. The vanishing of  ${{\left[ {1 - \left( {{{2GM} \mathord{\left/ {\vphantom {{2GM} r}} \right. \kern-\nulldelimiterspace} r}} \right)} \right]} \mathord{\left/ {\vphantom {{\left[ {1 - \left( {{{2GM} \mathord{\left/ {\vphantom {{2GM} r}} \right. \kern-\nulldelimiterspace} r}} \right)} \right]} {f^2 \left( {E\ell _p } \right)}}} \right. \kern-\nulldelimiterspace} {f^2 \left( {E\ell _p } \right)}}$ at point $r_H  = 2GM$  indicates the presence of an event horizon. Eq.~(\ref{eq2}) reproduces the standard static spherically symmetric black hole if  Eq.~(\ref{eq1+}) holds. According to Eq.~(\ref{eq2}), surface gravity on the event horizon is given by
\begin{align}
\label{eq3}
\kappa  =  - \frac{1}{2}\mathop {\lim }\limits_{r \to r_H } \sqrt { - \frac{{g^{11} }}{{g^{00} }}} \frac{{\left( {g^{00} } \right)^\prime  }}{{g^{00} }} = \frac{{g\left( {E\ell _p } \right)}}{{f\left( {E\ell _p } \right)}}\frac{1}{{4GM}},
\end{align}
where $\left( {g^{00} } \right)^\prime   = \partial _t \left( {g^{00} } \right)$. The above equation indicates that the surface gravity is modified by RG. It is well known that the Hawking temperature is defined in terms of surface gravity, that is, $T_H  = {\kappa  \mathord{\left/ {\vphantom {\kappa  {2\pi }}} \right. \kern-\nulldelimiterspace} {2\pi }}$. Using the Hawking temperature-surface gravity relation, the RG-corrected Hawking temperature can be expressed as
\begin{align}
\label{eq4}
T_H  = \frac{{g\left( {E\ell _p } \right)}}{{f\left( {E\ell _p } \right)}}\frac{1}{{8\pi GM}} = \frac{1}{{8\pi GM}}\left( {1 - \gamma E\ell _p } \right).
\end{align}
According to \cite{ch32,ch33,ch34,ch34a}, in the vicinity of the Schwarzschild surface, there is a relationship between the minimum value of position uncertainty $\Delta x$ of emitted particles and the Schwarzschild radius $r_H$, namely, $\Delta x\simeq r_H$. Then, by considering that the Heisenberg uncertainty principle $\Delta p\Delta x \ge 1$ still holds in RG, the energy uncertainty can be expressed as $\Delta E = {1 \mathord{\left/ {\vphantom {1 {\Delta x}}} \right.  \kern-\nulldelimiterspace} {\Delta x}}$. Therefore, one can obtain the following relation:
\begin{align}
\label{eq5}
E \ge \frac{1}{{\Delta x}} \simeq \frac{1}{{r_H }} = \frac{1}{{2GM}},
\end{align}
where $E$ is a lower bound on the energy of emitted particles. With the help of Eq.~(\ref{eq5}), the RG-corrected Hawking temperature can be rewritten as
\begin{align}
\label{eq6}
T_H  = T_0 \left( {1 - \frac{{\gamma \ell _p }}{{2GM}}} \right) = \frac{1}{{8\pi GM}}\left( {1 - \frac{{\gamma \ell _p }}{{2GM}}} \right),
\end{align}
where $T_0  = {1 \mathord{\left/ {\vphantom {1 {8\pi GM}}} \right. \kern-\nulldelimiterspace} {8\pi GM}}$  is the original Hawking temperature. Using the first law of black hole thermodynamics, the modified entropy is given by
\begin{align}
\label{eq7}
S & =  \int {T_H^{ - 1} dM} = 4 \pi GM^2 + 4\pi \gamma M\ell _p  + 2\frac{{\pi \gamma ^2 \ell _p^2 }}{G}{\rm{ln}}\left( {2GM - \gamma \ell _p } \right)
\nonumber \\
&  =  \frac{A}{{4\ell _p^2 }}\left[ {1 + 4\gamma \ell _p \sqrt {\frac{\pi }{A}}  + \frac{{8\pi \gamma ^2 \ell _p^2 }}{A}{\rm{ln}}\left( {\sqrt {\frac{\pi }{{4A}}}  - \gamma \ell _p } \right)} \right].
\end{align}
In the above expression, we use the area of the horizon  $A = 16\pi G^2 M^2$. It is clear that the modified entropy is related to the area of the horizon  $A$, Planck length $\ell _p$, and rainbow parameter  $\gamma$. Furthermore, there is a logarithmic correction term in Eq.~(\ref{eq7}), which is coincident with those results achieved in previous works (e.g., \cite{ch31+,ch32+}). When  $\gamma  = 0$, Eq.~(\ref{eq7}) reduces to the original entropy, namely, $S = {A \mathord{\left/ {\vphantom {A {4\ell _p^2 }}} \right. \kern-\nulldelimiterspace} {4\ell _p^2 }}$.

\section{The modified Newton's law of gravitation due to the rainbow gravity}
\label{sec3}
In the framework of quantum gravity, it is believed that information can be stored on a volume of space. This viewpoint is known as the holographic principle, which is supported by black hole physics and the anti-de Sitter/conformal field theory correspondence (AdS/CFT) correspondence. Now, following the idea of Verlinder, one can assume there is a screen in space and it can separate points. When the particles move from one side of the screen to the other side, the information is stored on the screen. In Ref.~\cite{ch13}, this screen is called a holographic screen. Therefore, according to the holographic principle and the second law of thermodynamics, when a test particle approaches a holographic screen, the entropic force of a gravitational system obeys the following relation:
\begin{align}
\label{eq8}
F\Delta x = T\Delta S,
\end{align}
where  $F$,  $T$, $\Delta S$, and $\Delta r$ represent the entropic force, the temperature, the change of entropy on holographic screen, and the displacement of the particle from the holographic screen, respectively. Eq.~(\ref{eq8}) means that a nonzero force leads to a nonzero temperature. Motivated by an argument in Ref.~\cite{ch1}, which posits that the change in entropy is related to the information on the boundary, that is $\Delta S = 2\pi m\Delta r$  with the mass of particle $m$. Hence, a linear relation between the  $\Delta S$ and  $\Delta r$ is given by
\begin{align}
\label{eq9}
\Delta S = 2\pi m\Delta r,
\end{align}
where  $\Delta r = {1 \mathord{\left/ {\vphantom {1 m}} \right. \kern-\nulldelimiterspace} m}$ with the mass of the elementary component. In Ref.~\cite{ch35}, Hooft demonstrated that the horizon of black holes can be considered as a storage device for information. Assuming that the amount of information is $N$  bits, and combining the equipartition law of energy with the holographic principle, the number of bits $N$  and area of horizon $A$ obey the following relation:
\begin{align}
\label{eq10}
N = \frac{A}{G}.
\end{align}
Substituting the entropy-area law $S = {A \mathord{\left/ {\vphantom {A {4G}}} \right. \kern-\nulldelimiterspace} {4G}}$ and modified entropy Eq.~(\ref{eq7}) into Eq.~(\ref{eq10}), the relation between $N$  and  $A$ can be rewritten as
\begin{align}
\label{eq11}
 N & = 4S = \frac{A}{{\ell _p^2 }}\left[ {1 + 4\gamma \ell _p \sqrt {\frac{\pi }{A}}  + \frac{{8\pi \gamma ^2 \ell _p^2 }}{A}{\rm{ln}}\left( {\sqrt {\frac{\pi }{{4A}}}  - \gamma \ell _p } \right)} \right].
\end{align}
In the above equation, we utilize the natural units, which says  $G = \ell _p^2$. Next, by adopting the equipartition law of energy, the total energy of the holographic system can be written as
\begin{align}
\label{eq12}
E = \frac{{NT}}{2}.
\end{align}
Now, substituting the relations  $E=M$ and $A = 4\pi r^2$, and Eq.~(\ref{eq8})-Eq.~(\ref{eq11}) into Eq.~(\ref{eq12}), the entropic force becomes
\begin{align}
\label{eq13}
 F & =  4\pi \frac{{Mm}}{N} =  F_0 \left[ {1 - \frac{{2\gamma }}{r}\ell _p  - \frac{{2\gamma ^2 }}{{r^2 }}{\rm{ln}}\left( {\frac{1}{{4r}} - \gamma \ell _p } \right)\ell _p^2 } \right]
 \nonumber \\
& = \frac{{GMm}}{{r^2 }}\left[ {1 - \frac{{2\gamma }}{r}\ell _p  - \frac{{2\gamma ^2 }}{{r^2 }}{\rm{ln}}\left( {\frac{1}{{4r}} - \gamma \ell _p } \right)\ell _p^2 } \right].
\end{align}
Obviously, Eq.~(\ref{eq13}) is the modified Newton's law of gravitation, and $F_0  = {{GMm} \mathord{\left/ {\vphantom {{GMm} {r^2 }}} \right. \kern-\nulldelimiterspace} {r^2 }}$ is the original Newton's law of gravitation. One can find that the modification is dependent not only on the mass of two bodies $M$ and $m$, the distances between the two bodies $r$, and Newton's gravitational constant  $G$ but also on the Planck length $\ell _p$  and the rainbow parameter  $\gamma$. It illustrates that the modified Newton's law of gravitation is valid at small scales due to the effect of rainbow gravity (or the effect of quantum gravity). Moreover, when  $\gamma=0$, Eq.~(\ref{eq13}) reduces to the conventional one.

\section{The modified Einstein's field equation due to the rainbow gravity}
\label{sec4}
In previous work, people found that Einstein's field equation can be derived from the entropic force approach. Therefore, in this section, we derive the modified Einstein's field equation via the RG-corrected entropic force. Taking into account Eq.~(\ref{eq11}), one can generalize the entropy-corrected relation to the following form:
\begin{align}
\label{eq14}
{\rm{d}}N = \frac{{{\rm{d}}A}}{G}\left[ {1 + 2\gamma \sqrt {\frac{{\pi G}}{A}}  - \frac{{4G\pi ^{\frac{3}{2}} \gamma ^2 }}{{A\left( {\sqrt \pi   - 2\gamma \sqrt {AG} } \right)}}} \right],
\end{align}
which is the bit density on a holographic screen. Next, consider a certain static mass configuration with total mass $M$ enclosed in the holographic screen  ${\cal S}$, and take the energy associated with it divided over $N$  bits. It is easy to find that each bit carries a mass equal to  ${T \mathord{\left/ {\vphantom {T 2}} \right. \kern-\nulldelimiterspace} 2}$. Hence, the total mass can be expressed as
\begin{align}
\label{eq15}
M = \frac{1}{2}\int_{\cal S} {T{d}N} ,
\end{align}
and the local temperature on the screen is
\begin{align}
\label{eq16}
T = \frac{{e^\phi  n^b \nabla _b \phi }}{{2\pi {\rm{ }}}},
\end{align}
where $e^\phi$  is the redshift factor as the local temperature $T$  is measured by an observer at infinity \cite{ch36}. Inserting the identifications for  ${\rm d N}$ and $T$, Eq.~(\ref{eq15}) can be rewritten as
\begin{align}
\label{eq17}
M & =  \frac{1}{{4\pi G}}\int_{\cal S} {e^\phi  \nabla \phi} \left[ {1 + 2\gamma \sqrt {\frac{{\pi G}}{A}}  - \frac{{4G\pi ^{\frac{3}{2}} \gamma ^2 }}{{A\left( {\sqrt \pi   - 2\gamma \sqrt {AG} } \right)}}} \right]. {d}A,
\end{align}
where the right hand side (RHS) of Eq.~(\ref{eq17}) is the modified Komar mass while the original Komar mass is $M_{\rm{K}}  = {1 \mathord{\left/ {\vphantom {1 {4\pi G}}} \right. \kern-\nulldelimiterspace} {4\pi G}}\int_{_{\cal S} } {e^\phi  \nabla \phi {d}A}$ \cite{ch13}.  Now, combining the Stokes theorem with the killing equation  $\nabla ^a \nabla _a  =  - R_a^b \xi ^a$, one can express the original Komar mass in terms of the Ricci tensor  $R_{ab}$ and the Killing vector $\xi ^a$ \cite{ch15,ch16}
\begin{align}
\label{eq18}
M_{\rm{K}}  = \frac{1}{{4\pi G}}\int_\Sigma  {R_{ab} n^a \xi ^b dV} .
\end{align}
Equating Eq.~(\ref{eq17}) and Eq.~(\ref{eq18}), one obtains
\begin{align}
\label{eq19}
M & = \frac{1}{{4\pi G}}\int_\Sigma  {R_{ab} n^a \xi ^b {d}V} + \frac{1}{{4\pi G}}e^\phi  \nabla \phi {\rm{ }}\int_\mathcal{S} \left[ {2\gamma \sqrt {\frac{{\pi G}}{A}}  - \frac{{4G\pi ^{\frac{3}{2}} \gamma ^2 }}{{A\left( {\sqrt \pi   - 2\gamma \sqrt {A G} } \right)}}} \right].{d}A.
\end{align}
In the above equation,  $\Sigma$  represents the three-dimensional volume bounded by the holographic screen and $n^a$  is the volume's normal. On the other hand, following the viewpoint that was proposed by Seiedgar, the total mass  $M$ can be expressed as the stress-energy $T_{\mu \nu }$ \cite{ch29}, that is,
\begin{align}
\label{eq20}
M = 2\int_\Sigma  {dV\left( {T_{\mu \nu }  - \frac{1}{2}Tg_{\mu \nu } } \right)} n^a \xi ^b.
\end{align}
Putting Eq.~(\ref{eq20}) into Eq.~(\ref{eq19}), one can obtain the following expression
\begin{align}
\label{eq21}
& \int_\Sigma {\left[ {R_{ab}  - 8\pi G\left( {T_{ab}  - \frac{1}{2}Tg_{ab} } \right)} \right]n^a \xi ^b dV}  = -  \int_\mathcal{S} {e^\phi  } \nabla \phi \left[ {2\gamma \sqrt {\frac{{\pi G}}{A}}  - \frac{{4G\pi ^{\frac{3}{2}} \gamma ^2 }}{{A\left( {\sqrt \pi   - 2\gamma \sqrt {AG} } \right)}}} \right]. dA.
\end{align}
The RHS of Eq.~(\ref{eq21}) is a correction term, which is caused by quantum gravity. In order to obtain Einstein's field equation, the RHS should be rewritten as \cite{ch319+}
\begin{align}
\label{eq22+}
- \int_{\cal S} {e^\phi  } \nabla \phi \left[ {2\gamma \sqrt {\frac{{\pi G}}{A}}  - \frac{{4G\pi ^{\frac{3}{2}} \gamma ^2 }}{{A\left( {\sqrt \pi   - 2\gamma \sqrt {AG} } \right)}}} \right].dA = -  GM\left[ {2\gamma \sqrt {\frac{{\pi G}}{A}}  - \frac{{4G\pi ^{\frac{3}{2}} \gamma ^2 }}{{A\left( {\sqrt \pi   - 2\gamma \sqrt {AG} } \right)}}} \right].
\end{align}
Substituting Eq.~(\ref{eq20}) and Eq.~(\ref{eq22+}) into Eq.~(\ref{eq21}), one has
\begin{align}
\label{eq23+}
\int_\Sigma  {\left\{ {R_{ab}  - 8\pi G\left( {T_{ab}  - \frac{1}{2}Tg_{ab} } \right) + 2G\left[ {2\gamma \sqrt {\frac{{\pi G}}{A}}  - \frac{{4G\pi ^{\frac{3}{2}} \gamma ^2 }}{{A\left( {\sqrt \pi   - 2\gamma \sqrt {AG} } \right)}}} \right]\left( {T_{ab}  - \frac{1}{2}Tg_{ab} } \right)} \right\}n^a \xi ^b dV}  = 0.
\end{align}
Finally, the deformed Einstein's field equation is
\begin{align}
\label{eq22}
R_{ab}  = 8\pi G\left( {T_{ab}  - \frac{1}{2}Tg_{ab} } \right)\left[ {1 + \frac{\gamma }{2}\sqrt {\frac{G}{{\pi A}}}  - \frac{{G\pi ^{\frac{1}{2}} \gamma ^2 }}{{A\left( {\sqrt \pi   - 2\gamma \sqrt {AG} } \right)}}} \right].
\end{align}
It is obvious that this field equation is affected by the geometry of spacetime, the energy-momentum tensor and the rainbow parameter $\gamma$. If $\gamma$  vanishes or the horizon area becomes too large, Eq.~(\ref{eq22}) reduces to the original Einstein's field equation.

\section{The modified Friedmann equation due to rainbow gravity}
\label{sec5}
In this section, according to the methods which were proposed by \cite{ch15,ch16,ch17,ch18,ch18x,ch25,ch25+,ch26+}, the RG corrected Friedmann equation is derived via the modified entropic force approach. First, let us consider a 4-dimensional rainbow FRW universe, whose linear element is given by
\begin{align}
\label{eq23}
{\rm d}{{s}}^2  = - \frac{{{\rm{d}}t^2 }}{{f^2 \left( {E\ell _p } \right)}} + \frac{{a\left( t \right)^2 }}{{g^2 \left( {E\ell _p } \right)}}\left( {\frac{{{\rm{d}}r^2 }}{{1 - kr^2 }} + r^2 {\rm{d}}\Omega ^2 } \right),
\end{align}
where ${\rm d}\Omega ^2  = {\rm d}\theta ^2  + \sin ^2 \theta {\rm d}\varphi ^2$ represents the metric of a two-dimensional unit sphere, $a$  is scale factor of the universe, and  $k$ is the spatial curvature constant. According to Refs.~\cite{ch40,ch41}, it is suitable to use the notion $\tilde r = {{f\left( {E\ell _p } \right)a\left( t \right)r} \mathord{\left/ {\vphantom {{f\left( {E\ell _p } \right)a\left( t \right)r} {g\left( {E\ell _p } \right)}}} \right. \kern-\nulldelimiterspace} {g\left( {E\ell _p } \right)}}$ and relation $h^{\mu \nu } \partial _\mu  \tilde r\partial _\nu  \tilde r = 0$; the modified dynamical apparent horizon of FRW spacetime becomes
\begin{align}
\label{eq24}
\tilde r_a = {1 \mathord{\left/
 {\vphantom {1 {\sqrt {H^2  + \frac{{f\left( {E\ell _p } \right)k}}{{g\left( {E\ell _p } \right)a^2 }}} }}} \right.
 \kern-\nulldelimiterspace} {\sqrt {H^2  + \frac{{f\left( {E\ell _p } \right)k}}{{g\left( {E\ell _p } \right)a^2 }}} }},
\end{align}
where $H = {{\dot a} \mathord{\left/ {\vphantom {{\dot a} a}} \right. \kern-\nulldelimiterspace} a}$  is the Hubble parameter with  $\dot a = \partial _t a$. Significantly, Eq.~({\ref{eq24}}) recovers the original apparent horizon of FRW spacetime when ${f\left( {E\ell _p } \right)}=1$ and ${g\left( {E\ell _p } \right)}=1$. If one assumes that the matter source in the FRW universe is a perfect fluid, the tress-energy tensor of this fluid can be expressed as
\begin{align}
\label{eq25}
T_{\mu \nu }  = \left( {\rho  + p} \right)u_\mu  u_\nu   + pg_{\mu \nu }.
\end{align}
Hence, the following continuity equation can be derived from the conservation law of energy-momentum
\begin{align}
\label{eq26}
\dot \rho  + 3H\left( {\rho  + p} \right) = 0.
\end{align}
For obtaining the modified Friedmann equation, we need to consider a compact spatial region with volume $V = \left( {{4 \mathord{\left/ {\vphantom {4 3}} \right. \kern-\nulldelimiterspace} 3}} \right)\pi \tilde r^3$. By incorporating the Newton's second law with the gravitational force, one has
\begin{align}
\label{eq27}
F = m\frac{{\partial ^2 \tilde r}}{{\partial t^2 }} = m\ddot ar\frac{{f\left( {E\ell _p } \right)}}{{g\left( {E\ell _p } \right)}}
 =  - G\frac{{Mm}}{{{\tilde r}^2 }}\chi \left( {\ell _p ,\gamma } \right),
\end{align}
where $\chi \left( {\ell _p ,\gamma } \right) = 1 - {{2\gamma \ell _p } \mathord{\left/ {\vphantom {{2\gamma \ell _p } {\tilde r}}} \right. \kern-\nulldelimiterspace} {\tilde r}} - {{2\gamma ^2 {\rm{ln}}\left( {{1 \mathord{\left/ {\vphantom {1 {4\tilde r}}} \right. \kern-\nulldelimiterspace} {4\tilde r}} - \gamma \ell _p } \right)\ell _p^2 } \mathord{\left/ {\vphantom {{2\gamma ^2 {\rm{ln}}\left( {{1 \mathord{\left/ {\vphantom {1 {4\tilde r}}} \right.  \kern-\nulldelimiterspace} {4\tilde r}} - \gamma \ell _p } \right)\ell _p^2 } {\tilde r^2 }}} \right. \kern-\nulldelimiterspace} {\tilde r^2 }}$. Substituting Eq.~(\ref{eq27}) into Eq.~(\ref{eq26}), one can obtain the following acceleration equation with some manipulations:
\begin{align}
\label{eq28}
\frac{{\ddot a}}{a} =  - \frac{4}{3}\pi G\rho \chi \left( {\ell _p ,\gamma } \right).
\end{align}
Furthermore, the total physical $M$ can be defined as
\begin{align}
\label{eq29}
M = \int_{\cal V} {dV\left( {T_{\mu \nu } u^\mu  u^\nu  } \right)}  = \frac{4}{3}\pi \tilde r^3 \rho ,
\end{align}
with energy density of the matter $\rho  = {M \mathord{\left/ {\vphantom {M V}} \right.\kern-\nulldelimiterspace} V}$, and the volume becomes $ V ={{4\pi \left[ {arf\left( {E\ell _p } \right)} \right]^3 } \mathord{\left/ {\vphantom {{4\pi \left[ {arf\left( {E\ell _p } \right)} \right]^3 } {3g^3 \left( {E\ell _p } \right)}}} \right. \kern-\nulldelimiterspace} {3g^3 \left( {E\ell _p } \right)}}$. However, in order to get the Friedmann equation, one should use the active gravitational mass  ${\cal M}={{4\pi \rho \left[ {a\tilde rf\left( {E\ell _p } \right)} \right]^3 } \mathord{\left/ {\vphantom {{4\pi \rho \left[ {a\tilde rf\left( {E\ell _p } \right)} \right]^3 } {3g^3 \left( {E\ell _p } \right)}}} \right. \kern-\nulldelimiterspace} {3g^3 \left( {E\ell _p } \right)}}$  to replace the total mass  $M$. Then, following the agreement of Refs.~\cite{ch16,ch18x}, the active gravitational mass is given by
\begin{align}
\label{eq30}
{\cal M}= &2\int_{\cal V} {\left( {T_{\mu \nu }  - \frac{1}{2}Tg_{\mu \nu } } \right)} u^\mu  u^\nu dV = \frac{4}{3}\left[ {\frac{{\pi \tilde rf\left( {E\ell _p } \right)}}{{g\left( {E\ell _p } \right)}}} \right]^3 \left( {\rho  + 3p} \right).
\end{align}
Next, using the active gravitational mass instead of the total mass in Eq.~(\ref{eq29}), the modified acceleration equation for the dynamical evolution of the FRW universe becomes
\begin{align}
\label{eq31}
\frac{{\ddot a}}{a} =  - \frac{4}{3}\pi G\left( {\rho  + 3p} \right)\chi \left( {\ell _p ,\gamma } \right).
\end{align}
It should be noted that the effect of correction terms in Eq.~(\ref{eq31}) is too small to detect because the apparent horizon radius ${\tilde r}$ is very large. However, according to the viewpoint in Ref.~\cite{ch26++}, the rainbow gravity parameter $\gamma$ can be investigated by using the modified acceleration equation for the dynamical evolution of the FRW universe.

As we know, prior research and observational data substantiates that our universe is accelerating \cite{ch42,ch43}. One of the possible explanations for this fact is that our universe is dominated by dark energy. Therefore, we can constrain the rainbow parameter $\gamma$ with the modified acceleration equation at the late time. When assuming our universe is dominated by the dark-energy, the modified acceleration equation at the late time satisfies the following expression:
\begin{align}
\label{eq31+}
\frac{{\ddot a}}{a} =  - \frac{4}{3}\pi G\left( {\rho  + 3p} \right)\left( {1 - \frac{{2\gamma \ell _p }}{{\tilde r}}} \right),
\end{align}
where $p = w \rho$. For the sake of discussion, we only retain the terms up to $\mathcal{O}(\gamma)$. Since dark energy has negative pressure, one has the relationship $1 + 3 w  < 0$. Hence, the upper bound of $\gamma$ is given by
\begin{align}
\label{eq31++}
\gamma  < {{\tilde r\ell _p } \mathord{\left/ {\vphantom {{\tilde r\ell _p } 2}} \right. \kern-\nulldelimiterspace} 2}.
\end{align}
Furthermore, when considering a universe dominated by dust, the modified acceleration equation at the late time is given by
\begin{align}
\label{eq32+}
\frac{{\ddot a}}{a} =   - \frac{4}{3}\pi G\rho \left( {1 - \frac{{2\gamma \ell _p }}{{\tilde r}}} \right).
\end{align}
In the equation above, we ignore the higher order terms of $\mathcal{O}(\gamma^2)$. For obtaining an accelerating universe, it is required that ${{\ddot a} \mathord{\left/ {\vphantom {{\ddot a} a}} \right. \kern-\nulldelimiterspace} a} > 0$. Therefore, the lower bound of the rainbow parameter turns out to be
\begin{align}
\label{eq32++}
\gamma  > {{\tilde r\ell _p } \mathord{\left/ {\vphantom {{\tilde r\ell _p } 2}} \right. \kern-\nulldelimiterspace} 2}.
\end{align}

Next, substituting the continuity equation Eq.~(\ref{eq26}) into the modified acceleration equation~(\ref{eq31}), and multiplying both sides with $\dot aa$, then integrating, one yields \cite{ch38,ch39}
\begin{align}
\label{eq32}
\frac{{d\left( {\dot a^2 } \right)}}{{dt}} = \frac{{8\pi G}}{3}{\frac{{d\left( {\rho a^2 } \right)}}{{dt}}} \chi \left( {\ell _p ,\gamma } \right).
\end{align}
Integrating both sides for each term of Eq.~(\ref{eq32}), the results is
\begin{align}
\label{eq33}
 H^2  + \frac{k}{{a^2 }} & = \frac{{8\pi G}}{3}\rho \left\{ {1 + \frac{1}{{\rho a^2 }}\int {\left[ { - \frac{{2\gamma }}{{ra}}\ell _p  - \frac{{2\gamma ^2 }}{{\left( {ra} \right)^2 }}} \right.} } \right. \left. { \times \left. {{\rm{ln}}\left( {\frac{1}{{4ra}} - \gamma \ell _p } \right)} \right]\ell _p^2 d\left( {\rho a^2 } \right)} \right\},
\end{align}
where $H$ means the Hubble parameter, $k$ is the spatial curvature. For $k=1$, $k=-1$, and $k=0$, one has a hyperspherical, hyperbolic, and flat FRW universe, respectively. According to the time-independent constant (or equation of state parameter)  $w  = {p \mathord{\left/ {\vphantom {p \rho }} \right. \kern-\nulldelimiterspace} \rho }$, the continuity equation Eq.~(\ref{eq29}) is integrated to give
\begin{align}
\label{eq34}
\rho  = \rho _0 a^{ - 3\left( {1 + w } \right)} ,
\end{align}
where  $\rho _0$ is an integration constant. By incorporating Eq.~(\ref{eq34}) with Eq.~(\ref{eq33}), then integrating, one has
\begin{align}
\label{eq36}
H^2  + \frac{k}{{a^2 }} = \frac{{8\pi G}}{3}\rho \left\{ {1 + \left( {1 + 3w} \right)\left\{ { - \frac{{2\gamma }}{{\tilde r\left( {2 + 3w} \right)}}\ell _p } \right.} \right. \left. {\left. { + \frac{{2\gamma ^2 }}{{9\tilde r^2 \left( {1 + w} \right)^2 }}\left[ {1 - 3\left( {1 + 2} \right){\rm{ln}}\left( {\frac{1}{{4\tilde r}} - \ell _p \gamma } \right)} \right]\ell _p^2 } \right\}} \right\},
\end{align}
where ${\mathcal{O}\left( {\gamma^3 ,\ell _p ^3} \right)}$ is the higher order correction terms. With the help of the expression of dynamical apparent horizon $\tilde{r}_a$, Eq.~(\ref{eq36}) can be rewritten as
\begin{align}
\label{eq37}
& \left( {H^2  + \frac{k}{{a^2 }}} \right)\left\{ {1 - \left( {1 + 3w } \right)\left\{ { - \frac{{2\gamma }}{{\left( {2 + 3w } \right)}}\sqrt {H^2  + \frac{{k^2 }}{{a^2 }}} \ell _p } \right.} \right.
    \nonumber \\
&  + \frac{{2\gamma ^2 }}{{9\left( {1 + w } \right)^2 }}\left( {H^2  + \frac{{k^2 }}{{a^2 }}} \right)\left[ {1 - 3\left( {1 + w } \right)} \right.
     \nonumber \\
& \left. { \times \left. {\left. {{\rm{ln}}\left( {\frac{1}{4}\sqrt {H^2  + \frac{{k^2 }}{{a^2 }}}  - \ell _p \gamma } \right)\ell _p^2 } \right]} \right\}} \right\} = \frac{{8\pi G}}{3}\rho.
\end{align}
Finally, we derive the modified Friedmann equation from the RG corrected entropic force. Eq.~(\ref{eq37}) shows that the deformed Friedmann equation is related not only to the spatial curvature  $k$, Hubble parameter  $H$, scale factor of the universe $a$, and equation of state parameter $\omega$ but also to the rainbow parameter  $\gamma$ and Planck length $\ell_p$. It means that the Friedmann equation is observer-dependent. Due to the effect of rainbow gravity, the deformed Friedmann equation~(\ref{eq37}) becomes susceptible when the apparent horizon approaches the order of Planck scale. Therefore, our result can be used to analyze the early stage of the universe. In addition, when the apparent horizon radius becomes too large or  $\gamma=0$, Eq.~(\ref{eq37}) recovers the original Friedmann equation.

\section{Conclusions}
\label{sec6}
In the paper, we studied the RG-corrected entropic force. First, we investigated the RG-corrected thermodynamics, and found that RG leads to a new mass-entropy relation. Then, following Verlinde's new perspective on the relation between gravity and thermodynamics, the quantum corrections to the entropic force and Newton's law of gravitation are obtained. Moreover, we further extended our study to the relativistic case as well as cosmological case; hence, we derived the modified Einstein's field equation and the modified Friedmann equation via the RG-corrected entropic force. Our results  reveal that the modified Newton's law of gravitation, modified Einstein's field equation, and the modified Friedmann equation are influenced by the effect of rainbow gravity (rainbow parameter $\gamma$). In other words, the results are observer-dependent. Meanwhile, it has been found that the results are sensitive to a scale that is comparable to the Planck energy scale. However, when rainbow parameter $\gamma$ vanishes or the scale becomes too large, the modifications reduce to the original cases. Therefore, one can use the modified Newton's law of gravitation and Einstein's field equation to investigate the gravitational interaction between objects at the sub-$\mu m$ scale, and analyze the properties of the early universe by using the modified Friedmann equation. Moreover, according to the observational data of today's astronomy, one can also constrain the rainbow parameter $\gamma$ by various energy combinations of the universe. Furthermore, it is easily found that our results are different from those in Ref.~\cite{ch18x} since we adopt a new RG model, which was proposed by Magueijo and Smolin. Recently, the authors in Ref.~\cite{ch43} show that velocity of light in the early universe was faster than it is now, and RG model we used in this paper just fits their view point. Therefore, we think our results are more suitable for the study of the physical properties of the early universe.

\end{document}